\documentclass{article}

\usepackage{arxiv}

\usepackage[utf8]{inputenc} 
\usepackage[T1]{fontenc}    
\usepackage{hyperref}       
\usepackage{url}            
\usepackage{booktabs}       
\usepackage{amsfonts}       
\usepackage{nicefrac}       
\usepackage{microtype}      
\usepackage{lipsum}

\title{A position paper on GDPR compliance in sharded blockchains: rehash of old ideas or new interesting challenges?}

\author{
  Narasimha Raghavan Veeraragavan \\
  Department of Informatics\\
  University of Oslo\\
  Norway \\
  \texttt{raghavan@ifi.uio.no} \\
   \And
Kaiwen Zhang\\
  Department of Sofware and IT Engineering\\
 \`{E}TS Montr\`{e}al\\
  University of Qu\`{e}bec \\
  \texttt{kaiwen.zhang@etsmtl.ca} \\
}

\begin{document}
\maketitle

\begin{abstract}
Sharding has emerged as one of the common techniques to address the scalability problems of blockchain systems. To this end, various sharding techniques for blockchain systems have been proposed in the literature. When sharded blockchains process personal data, the data controllers and the data processors associated with the sharded blockchains need to be compliant with the General Data Protection Regulation (GDPR).  To this end, this article makes the first attempt to address the following key question: to what extent the existing techniques developed by different communities such as the distributed computing community, the distributed systems community, the database community, identity and access control community and the dependability community can be used by the data controllers and data processors for complying with the GDPR requirements of data subject rights in sharded blockchains? As part of answering this question, this article argues that there is a need for cross-disciplinary research towards finding optimal solutions for implementing the data subject rights in sharded blockchains.
\end{abstract}


\section{Introduction}
With the advent of Bitcoin~\cite{bitcoin} and Ethereum~\cite{ethereum} and Fabric~\cite{fabric}, the blockchain technology has gained significant attention in industry, academia, and government. One of the fundamental problems of blockchain technology is about the lack of ability to scale regarding transaction throughput. To this end, sharding has emerged as a promising technique to address the scalability concerns of blockchain technology. The concept of sharding refers to the process of dividing the computation, communication, and storage resources of a blockchain network into multiple shards. Examples of sharded blockchains include RapidChain~\cite{rapid}, Elastico~\cite{elastico}, Omniledger~\cite{omni}, ChainSpace~\cite{chain}, RSCoin~\cite{rs}, SharPer~\cite{sharper}. 

The EU members states designed General Data Protection Regulations (GDPR)~\cite{gdpr} with the goals of harmonizing the data privacy laws across all EU member countries and offering greater protection and rights to the individuals. GDPR introduced 99 articles covering the key principles around how organizations should process personal data of the EU citizens and these articles have been effective since 25 May 2018. 

As per the GDPR, when an organization collects personal data of EU citizens (or data subjects), the organization can play the role of a data controller when the organization determines the means and purposes of processing. On the other hand, an organization can play the role of data processor when the organization performs personal data processing as per the instruction of the data controller~\cite{gdpr}. 

When sharded blockchains store and process personal data (as defined in the Article 4 of GDPR~\cite{gdpr}), the data controller associated with the sharded blockchain systems need to be compliant with regulations such as GDPR. 

%
%

GDPR has allocated certain rights to the end-users that can be exercised against the data controllers that process the personal data of the users. At the essential level, GDPR articles from 12 to 22 describe all the data subject rights. One of the important rights is the ``Right of Access'' (Article 15 in~\cite{gdpr}), where the data controller is obligated to provide a copy of all personal data available in a sharded blockchain to the corresponding legitimate data subject. Successful execution of the ``Right of Access'' may help the user to exercise further rights such as the ``Right to Rectify''~(Article 16 in~\cite{gdpr}) and the ``Right to be Forgotten''(Article 17 in~\cite{gdpr}) since the user will have complete knowledge about his/her personal data.

Although the research community has begun tackling the GDPR challenges in blockchain~\cite{gdprb}~\cite{berasure}~\cite{gdprp}, an often asked question is: what are the new technical challenges for sharded blockchain systems towards data subject rights, and what problems can we solve using the existing techniques. To this end, this article makes the first attempt to answer the following central question:
\textbf{can the ``Right of Access'' be honored through existing techniques in the context of sharded blockchains where the storage resources are sharded?} 

Towards answering the above question, the following are the list of contributions made by this article:
\begin{itemize}
\item We analyze what existing principles and techniques from different communities can be exploited by the data controllers for exercising the right of access in sharded blockchains. 
\item We introduce several open cross-disciplinary research questions for each step involved in exercising the right of access. These questions promote cross-disciplinary research activities towards enabling the data controllers of sharded blockchains to demonstrate stronger GDPR compliance. 
\end{itemize}

The rest of this paper is organized as follows: Section~\ref{background} discusses the background and related work. Section~\ref{identity} discuss the principles, techniques and open challenges involved in verifying the identity of the data subject by the data controller. Additionally, Section~\ref{find} provides the principles, techniques and open challenges in finding the appropriate nodes that host the personal data of the data subject request issuer. Furthermore, Section~\ref{DGS} discusses the principles, techniques and open challenges involved in gathering the relevant data from multiple nodes and make the gathered data available to the data subject request issuer. Finally, Section~\ref{con} provides the conclusion. 
\section{Background and related work}\label{background}
\subsection{General Design of Sharded Blockchains}
The typical design of all sharded blockchains include six major components~\cite{soks}:
\begin{itemize}
 \item Identity establishment. Each node has to get an identity to participate in the consensus protocol of sharded blockchains.~\cite{soks} provides an overview of different identity establishment algorithms for nodes in various sharded blockchains. 
 \item Committee (shard) formation.  After a node gets an identity, the node needs to be assigned to a committee. There are different node-to-committee assignment algorithms~\cite{soks} exist in the literature striking tradeoffs between security and performance.
 \item Overlay setup for committee. As soon as the committees are created, each node in the committee should discover other nodes belonging to the same committee. Various node discovery protocols~\cite{comsat} from peer to peer research work is used here. 
 \item Intra-committee consensus. After the overlay setup is completed, whenever a transaction gets submitted to a node, the node triggers consensus protocol within the members of the node's committee to agree on the transaction. 
 \item Inter-committee consensus. After the agreement on transaction is made with the local context through intra-committee consensus, a global agreement is required with other committees towards creating a unified view of the ledger across all shards. Several variants of PoX and BFT based research works are typically used for implementing the intra and inter committee consensus protocols~\cite{soks}~\cite{consensus}.   
 \item Epoch reconfiguration. The nodes are shuffled among committees periodically for security reasons. In other words, the nodes will start executing the node-to-committee assignment algorithms periodically to guard the system against an adaptive attacker~\cite{soks}. 
 
\end{itemize}
In state-of-the-art literature, there are several sharded blockchains proposed with various architecture goals. The differences in the goals lead to the differences in the choices of algorithms for implementing each of the above components in sharded blockchains.~\cite{soks} and~\cite{shardings} provide surveys of architectural comparison of several sharded blockchains. 

\subsection{The case of GDPR in Sharded Blockchain}
The examples of personal data processed through sharded blockchain systems are listed as follows: 

The first example is the public key of a user, which serves as an identifier, as mentioned in the Recital 30 GDPR~\cite{gdpr}. All the transactions of a user in blockchain systems are linked to the public key owned by the user. The combination of public key with additional identifiers has the potential to reveal the natural person involved with the transactions. For example, Trade Exchange platforms are required to link the public key with real-world identities to comply with other regulations such as Anti-Money Laundering (AML) and Know Your Customer (KYC). Furthermore, users can voluntarily link the public key to the real-world identity to send and receive donations. In these cases, the public key is considered as personal data~\cite{gdprb}.  

Another interesting example is the payload in transactions that contain personal data.  For example, a university diploma issued through the blockchain system shall have information about the name and date of birth of the candidate in the payload of the transactions that issue the diploma to the candidate~\cite{mitb}. 

In the state-of-the-art literature, several works~\cite{gdprb}~\cite{berasure}~\cite{gdprp} address the GDPR challenges in the blockchain. First, some works~\cite{off1}~\cite{off2} argue that personal data should not be stored in the on-chain, and only the link to the personal data along with the time stamp and a hash of the outsourced data should be stored in the on-chain for verification. As a result, these works assume that the data removal or executing the “right to be forgotten” becomes easier for the data controllers and data processors since they do not need to deal with the immutability property of blockchains. 

We argue that the off-chain storage solutions for the right to be forgotten is achievable only if the off-chain storage of personal data has a trusted peer-to-peer network providing the storage and deletion services for personal data. For example, if the personal data is stored in the untrusted peer-to-peer storage network such as Inter Planetary File System (IPFS)~\cite{ipfs}, there could be non-active nodes or nodes that can simply ignore to perform the delete command leading to non-compliance. Additionally, from the absolute approach view point~\cite{enc}, the hashed data on the blockchain are perceived as pseudonymous data and not fully anonymous data. Hence the hashed data are subjected to GDPR, and appropriate protection mechanisms are required to protect the hashed data. 

On the other hand, we argue that it is easier to execute the right of access and verify the results of the right of access when the personal data is stored in the off-chain storage such as IPFS. The core storage principles of IPFS, such as unique identification via content addressing, content linking via directed acyclic graphs, and the content discovery via distributed hash tables, make the task of retrieving and verifying the content (personal data) associated with the request relatively easier under the assumption that the on-chain stores the complete list of content identifiers associated with the request of the data subject. 

Another interesting research direction is to store all the personal data only in the on-chain and use encryption techniques to comply with the GDPR requirements such as Article 32 in~\cite{gdpr}. The original ideas behind storing encrypted data in the on-chain are to achieve high confidentiality and comply with the right to be forgotten where the decryption keys can be deleted by the data controller/processors under the assumption that the inaccessibility to data is considered as deletion of data. However, the French data protection authorities disagree; the inaccessible encrypted personal data is not equal to erasure and residual risks to data subjects must be considered into account for further protection assessments~\cite{cnil}. 

Furthermore, the data controllers/processors have the burden of either managing the decryption keys of many users or securely processing the decryption keys managed by the end-users towards getting access to the data. Additionally, the encryption/decryption process may have an impact on the performance of the blockchain systems. Hence, we argue that storing the personal data in the encrypted form in the blockchains further complicates the execution of data subject rights such as the right to be forgotten and the right of access.  


We believe that in non-sharded blockchains, right to be forgotten is the most profound data subject rights to be honored by the data controllers/processors for the above-mentioned reasons. Right to access is relatively easier to execute since a full node in the blockchain host the entire copy of the ledger, the search of personal data associated with the public key requires scanning the ledger of one node. 

However, the personal data associated with a public key is distributed randomly across several nodes in sharded blockchains. As a result, the right to be forgotten and right of access share a common challenge, such as finding the target nodes that host the personal data related to the public key. After finding the target nodes, exercising the right of access has more unique challenges such as data gathering and data storage in sharded blockchain environment in compared to the non-sharded blokchains. On the other hand, exercising the right to be forgotten have similar concerns in both sharded and non-sharded blockchain environments after identifying the target nodes.

\subsection{Steps in exercising the Right of Access}

The following are the six fundamental steps that data controllers of sharded blockchains may follow towards honoring the ``Right of Access'' request from a data subject. These steps are identified based on the practices from industry~\cite{me}{}.  
\begin{itemize}
\item Verifying the identity of the user (data subject).
\item Finding the appropriate nodes that host the personal data related to the public key.
\item Data gathering from only selective nodes such that all distinct personal data elements  linked to the public key in the system is retrieved and redundant data elements are avoided.
\item Generating unified structured view of gathered data such that the data subjects has meaningful and transparent understanding about the personal data processed in the system.  
\item Securely delivering the structured view to the data subject who exercised the right.   
\item Delete any ephemeral storage of structural view and all personal data related to the public key as soon as the delivery of the structured view  is successful. 
\end{itemize}

\section{Verifying the identity} \label{identity}
The first step in exercising the right of access or for any other data subject rights is to verify if the issuer of the request is a legitimate issuer of the request.  

The proportionality principle~\cite{propl} from a security perspective states that the effort required for the verification process should be proportional to the nature of personal data that is being protected by the sharded blockchain systems. Furthermore, NIST SP 800-63-A~\cite{nist} addresses different identity proofing requirements varying from no association with specific real-life identity to the physical presence of the data subject for different (very low to very high) risk profiles. To this end, the choice of the authentication scheme should depend on the risk profile of the data subjects.
According several research works have conducted on risk based adaptive authentication techniques~\cite{auth}~\cite{auth2}. 

When a data controller of sharded blockchain services collect and process different categories of personal data of data subjects varying from demographic data to health data, the amount of personal data collected per data subject and the sensitivity of the data category strongly influences the risk profile of that data subject. Hence, a same data controller may need to adapt the authentication schemes for different risk profiles of data subjects.

In addition to the user identity required for data subject authentication, a node identity is required for nodes to participate in the consensus protocol of sharded blockchains as discussed in Section~\ref{background}.  A user identity should be directly or indirectly related to the node identities in order to exercise the data subject rights (for example to fetch or delete or correct the personal data associated with the user identity) in sharded blockchains. However, we believe establishing this relationship in a secure manner without compromising performance on exercising the data subject rights is not trivial even with minority of untrusted nodes in peer to peer environments. 

In essence, there are several disjoint research works exist on the human-identification (authentication) protocols~\cite{hauth1}~\cite{hauth2}, peer-to-peer authentication protocols~\cite{mauth1} and privacy risk assessments~\cite{priam}~\cite{r2} and architectural styles of sharded blockchains~\cite{soks}.  These disjoint research works do not help us to understand the design of suitable risk-profile dependent end to end authentication workflow in sharded blockchains. To this end, there is a need for scientific research to systematically unify the research works in these four areas towards answering the following question: 
\begin{itemize}
\item How the establishment of user identities should influence the establishment of machine (node) identities for various risk profiles of data subjects in different architectural styles of sharded blockchains?
\end{itemize}
Answering the above question will provide guidelines to the architects, application, and infrastructure developers towards provisioning appropriate identity and access management schemes for the users and machines in different architectural styles of sharded blockchains.  

Furthermore, there are many identity and access management projects based on blockchains~\cite{alastria}~\cite{uport}~\cite{sovrin}~\cite{indy} are emerging towards ``Self-Soverign Identity'' and compliance with GDPR. We believe the key challenge associated with all these projects is interoperability with different architectural styles of sharded blockchains. And, the above mentioned research question should be investigated in each of the projects. 
\section{Finding the appropriate nodes and data}\label{find}
After the verification, the next step in the process of honoring the ``Right of Access''  is to examine whether or not any personal data about the authenticated data subject request issuer is available in the sharded blockchain system. If there are no personal data exists in a sharded blockchain, then the request of the data subject can be responded immediately. Else, the personal data processed in the sharded blockchain should be delivered to the legitimate data subject request issuer. 

To this end, the following questions need to be addressed:
\begin{itemize}
\item Are there any nodes that host the personal data linked to the public key of the data subject in the sharded blockchains? 

\item How to keep track of the list of nodes that store personal data elements belonging to a given public address associated with the user under the following constraints: reconfiguration under epochs, and handling churns? 

\item How should nodes organize data such that the retrieval of data belonging to individual users is fast and secure from a single node without compromising the security and performance guarantees offered natively by the sharded blockchain architecture? 

\item In sharded blockchain architecture, multiple nodes can host the replicated data, retrieving the same data from various nodes would result in wastage of communication and processing cost. To this end, how the data can be retrieved from only one out of many replicas? 
\end{itemize}

The above questions can be seen as the problems of efficient search/location of data items and routing architecture in sharded database and peer-to-peer technologies. In this case, the lookup key for search operation is the public key available in the data subject request. And, the corresponding value is the personal data items linked to that address. 

Key-value lookups is a very well known problem in sharded databases and several solutions exists to address this problem~\cite{nosql}. However, the direct applicability of these techniques to the sharded blockchains require to understand the key differences in the properties between the sharded database and blockchains. 

Sharding techniques in database systems assume crash-failure model for the nodes involved in shards; a faulty node does not reply to the query in a shard of a database. In contrast, most of the sharded blockchains assume byzantine-failure models, where the faulty nodes can provide arbitrary responses. As a consequence, the security requirements are more stricter for designing consensus protocol in large scale public deployments of sharded blockchains. 

Furthermore, sharded databases are developed on top of the trusted infrastructure with pre-established identities~\cite{nosql}.  In contrast, sharded blockchains like Elastico~\cite{soks}  are designed to work on trust-less infrastructure with no pre-established identities. To this end, we do not think the the key-value lookup techniques from sharded databases research are directly applicable to certain architecture styles of sharded blockchains. 

On the other hand, if some sharded blockchains assume crash-fault model and deployed in trusted infrastructure, then we believe that key-value lookup techniques developed in sharded database research can be directly applicable in these sharded blockchains. Nevertheless, further research is required to investigate the overhead introduced by these techniques when directly applied in sharded blockchains. 

Several research papers in peer-to-peer structured  and unstructured overlay schemes have already provided solutions to key-value lookups under network churn and byzantine models~\cite{comsat} . Each peer-to-peer solution has its tradeoff in terms of routing complexity, reliability, security and communication overhead~\cite{comsat}. Due to these tradeoffs, the existing peer-to-peer protocols may not be directly suitable to solve the key-value lookup problems in certain sharded blockchains with byzantine nodes and trust-less infrastructure. To this end, we think the following research questions should be addressed: 
\begin{itemize}
\item To what extent the existing P2P overlay approaches can be used towards designing secure key-value protocols for sharded blockhains with byzantine nodes and decentralized infrastructure?
\item Is it possible to build a decentralized and secure adaptive/reflective middleware that can provide search functionality by adapting its internal structure and behavior for different architectures of sharded blockchains? 
\end{itemize}

\section{Data Gathering and Storage} \label{DGS}
The next step in the process of exercising the data subject rights is to gather the data from multiple nodes that host data elements linked to the public key involved in the data subject request. 

To this end, the following question needs to be addressed: 

How to effectively gather and store data from different nodes that host the personal data linked to the public key of the data subject for the following scenarios:
\begin{itemize}
\item A lower number of nodes containing a lower volume of data
\item A higher number of nodes containing a lower volume of data
\item A higher number of nodes containing a higher volume of data
\item A lower number of nodes containing a lower volume of data 
\end{itemize}
Peer-to-peer research work in the context of data aggregation may partially help in answering the question. The data aggregation problem in peer-to-peer is about calculating the global properties of the system derived by the distributed computation functions like COUNT, AVERAGE, and SUM on peer-to-peer nodes~\cite{agg}. These algorithms have different tradeoffs in terms of accuracy, space, and time complexity. Some of these algorithms can be used as an input to determine the storage and network capacity requirements for gathering and downloading the data associated with per public key. These inputs provide valuable information to define the storage architecture required for delivering the gathered data to the data subject. 

Furthermore, considering the fact that different data elements are linked to different public keys, and each public key corresponds to a user,  we believe that the storage architecture should partition the storage unit per public key and use that storage unit to store the all the data linked to a single public key at the time of exercising the data subject request. We believe that the partitioning the storage unit per public key allows to define fine granular authentication and authorization scheme to allow only the data subject request issuer to retrieve the data from the storage unit. Furthermore, the storage unit should have the capability to support the deletion of gathered data since there is no purpose in storing the data after the user has downloaded the target data. 

To promote transparency and meaningful understanding about the personal data processed in the system, the data controller shall present the data in a structured manner to the data subject request issuer. We believe that when a higher volume of data exist then Big Data storage architectures~\cite{bigs} and Big Data analytics techniques~\cite{biga} can be used to gather, store and deliver the personal data associated with the public keys. 

The following are the interesting research questions that need to be addressed in this step. 
\begin{itemize}
\item How to verify in a trustworthy manner that the data controller has provided all the personal data linked with the public keys in sharded blockchains? 
\item How to make sure that the data controller has not disclosed the personal data associated with the public keys to unauthorized entities? 
\end{itemize} 
\section{Conclusion} \label{con}
We believe that the satisfaction of the data subjects with respect to the data subject rights related services offered by the data controller is a key to build trust relationships between the data subjects and the data controller. To this end, last but not least open research question is as follows: how to measure the Quality of Experience (QoE)~\cite{qoe1}~\cite{qoe2} perceived by the data subjects when the data controllers of sharded blockchains offer the data subject rights related services to the data subjects especially in presence of failures~\cite{qoe3}?  

In summary, we believe there are several open research questions that require inter-disciplinary research to find optimal solutions for the technical problems of GDPR compliance for various architectural styles of sharded blockchains. To this end, this article aims to act as a first catalyst to promote the complex inter-disciplinary research in GDPR compliance for sharded blockchains.

\bibliographystyle{unsrt}  
\bibliography{references}  


\end{document}